\documentclass[aps,prb,twocolumn,showpacs]{revtex4}
\newcommand{\bsim}{\mbox{\raisebox{-0.1cm}{$\;
\stackrel{\textstyle>}{\sim}\;$}}}
\newcommand{\lsim}{\mbox{\raisebox{-0.1cm}{$\;
\stackrel{\textstyle<}{\sim}\;$}}}
\usepackage{psfig}

\begin{document}

\title{Small Fermi energy, zero point fluctuations and nonadiabaticity
in MgB$_2$}

\author{L. Boeri$^{1,2}$, E. Cappelluti$^{2,1}$,
and L. Pietronero$^{1,2}$}

\affiliation{$^1$Dipart. di Fisica, Universit\`{a} di Roma ``La Sapienza",
P.le A. Moro, 2, 00185 Roma, Italy}

\affiliation{$^2$INFM and ``Istituto dei Sistemi Complessi'' del CNR,
v. dei Taurini 19, 00185 Roma, Italy}

\date{\today}

\begin{abstract}
Small Fermi energy effects are induced in MgB$_2$ by the low hole
doping in the $\sigma$ bands which are characterized by a Fermi energy
$E_{\rm F}^\sigma \sim 0.5$ eV. We show that, due to the particularly
strong deformation potential relative to the $E_{2g}$ phonon mode,
lattice fluctuations are reflected in strong fluctuations in the
electronic band structure.  Quantum fluctuations associated to the
zero-point lattice motion are responsible for an uncertainty of the
Fermi energy of the order of the Fermi energy itself, leading to the
breakdown of the adiabatic principle underlying the Born-Oppenheimer
approximation in MgB$_2$ even if $\omega_{\rm ph}/E_{\rm F} \sim
0.1-0.2$, where $\omega_{\rm ph}$ are the characteristic phonon
frequencies.  This amounts to a new nonadiabatic regime, which could
be relevant to other unconventional superconductors.
\end{abstract}
\maketitle

Four years after the discovery of superconductivity at 39 K in
MgB$_2$,\cite{akimitsu} the superconducting properties of this
material are still object of investigation.
One of the anomalous characteristics of MgB$_2$ with respect to
conventional ME superconductors is the remarkable
smallness of the Fermi energy $E_{\rm F}^\sigma$
associated with the hole-like $\sigma$ bands,
which are the most involved in the Cooper pairing.
ME theory holds true indeed
only if the Fermi energy $E_{\rm F}$ is much larger than any other
energy scale of the system.  Several analyses indicate however that
this is not the case of MgB$_2$. LDA calculations, for instance,
estimate the energy distance between the chemical potential and the
top of the $\sigma$ bands to be $\sim$ 0.4-0.6 eV,\cite{an,kortus}
in agreement with ARPES measurements.\cite{uchiyama}
Estimates of the Fermi energy can be inferred also
from penetration depth measurements giving
$E_{\rm F} = 3200$ K in MgB$_2$,\cite{papagelis}
and even less in Al and C doped
compounds.\cite{papagelis,serventi}

In previous studies we have analyzed some of the effects
related to a small Fermi energy $E_{\rm F}^\sigma$
on different electronic and vibrational properties of MgB$_2$.
One of these effects is the breakdown of Migdal's theorem,
and consequently of the Migdal-Eliashberg diagrammatic theory, which occurs
when the Fermi energy
$E_{\rm F}^\sigma$ and the phonon energy $\omega_{\rm ph}$ are comparable 
($\omega_{\rm ph}/E_{\rm F}^\sigma \lsim 1$).\cite{ccgps,ccgp}
Another one is the remarkable
anharmonicity of the $E_{2g}$ phonon mode, 
related to the
splitting in energy of the $\sigma$ bands under the $E_{2g}$
distortion.\cite{bbcp}

In this paper we want to discuss in greater detail new physical
consequences of the comparable size of $E_{\rm F}^\sigma$ and the
$E_{2g}$ splitting energy of the $\sigma$ bands in MgB$_2$.
In particular we review in a critical way
the validity of the adiabatic Born-Oppenheimer (BO) approximation for
the $E_{2g}$ phonon, which is the most relevant
to the superconducting pairing.  As our
main result we show that, due to the large quantum fluctuations
associated with the zero point motion,
the BO principle is broken down {\em independently} of the
$\omega_{\rm ph}/E_{\rm F}^\sigma$ ratio.
This novel kind of adiabatic breakdown
is thus shown to be related to the parameter
$\kappa = g_{E_{2g}}/E_{\rm F}^\sigma$, where $g_{E_{2g}}$
is the electron-phonon matrix element which couples the electrons
in $\sigma$ bands to the $E_{2g}$ phonon mode in MgB$_2$.
A simple quantum analysis gives $\kappa \sim 0.9-1$,
strongly questioning the reliability of a Born-Oppenheimer
based analysis.

The Born-Oppenheimer principle is implicitly assumed
in writing down an effective electron-phonon model starting from 
{\em ab-initio} techniques.
In simple terms, the BO principle permits to describe the
full many-body wave function
$\Psi({\bf r};{\bf u})$ in terms of two separate
quantum systems: a purely electronic problem, which depends
parametrically on the ion variables, and an effective lattice problem,
where the electronic degrees of freedom have been
integrated out.\cite{grimvall}
The total wave function
$\Psi({\bf r};{\bf u})$ can thus be written as the product of two
``partial'' wave functions:
\begin{equation}
\Psi_{\alpha,n}({\bf r};{\bf u}) \simeq \chi_{\alpha,n}({\bf
u}) \varphi_n({\bf r};[u]),
\label{bo}
\end{equation}
where ${\bf r}$ and ${\bf u}$ represent
the electronic and lattice degrees of freedom respectively and
$(\ldots;[u])$ indicates a parametric (not quantum) dependence on the
lattice variable.
The index $n$ identifies the quantum number
of the electronic state and the label $\alpha$ the phonon eigenstate
in the effective lattice potential which depends on the electronic cloud
$\varphi_n({\bf r};[u])$.

In the spirit of the BO approximation,
the electronic ground state wave function $\varphi({\bf r};[u])$
is obtained as the lowest energy electronic state available for fixed $u$.
M. Born and J.R. Oppenheimer showed that
Eq. (\ref{bo}) can be considered a good approximation
as far as $(-\hbar^2 /M) \nabla_u^2 \varphi({\bf r};[u])$ 
($M$ being the atomic mass) is
negligible.
This assumption corresponds to the adiabatic
hypothesis that the electronic dynamics of $\varphi({\bf r};[u])$,
is much faster than the lattice one.
In common metals 
a widely used parameter to evaluate this condition is the
adiabatic ratio $\omega_{\rm ph}/E_{\rm F}$, where $E_{\rm F}$
can be estimated in the undistorted $u=0$ ground state configuration,
$E_{\rm F}=E_{\rm F}[u=0]$,
and $\omega_{\rm ph}$ is the lowest energy excitation of the
$\chi_{\alpha,n}$ phonon spectrum.
However it should be noted that, in order for the adiabatic principle
to apply, the condition
$(-\hbar^2 /M) \nabla_u^2 \varphi({\bf r};[u]) \sim 0$
must be fulfilled in {\em all the region} of the {\bf u}-space 
where $\chi({\bf u})$ has a sizable weight.
This constraint results to be much stronger than
the conventional adiabatic condition, 
$\omega_{\rm ph}/E_{\rm F} \ll 1$,
since it implies the adiabatic ratio to be negligible for
generic lattice distortions $u$,
namely $\omega_{\rm ph}/E_{\rm F}[u] \ll 1$,
where the lattice wave-function $|\chi({\bf u})|^2$ provides
the range of the relevant $u$-space.
As we are going
to see, the breakdown of this assumption for a relevant range of the
$E_{2g}$ lattice distortion $u$ is expected to be
responsible for the partial failure of the
BO approximation in MgB$_2$. 

In order to test the validity of the BO approximation,
the example of frozen phonon techniques is particularly instructive
since they are a direct implementation of the BO
principle. In the frozen-phonon approach the electronic
structure and the total energy of the system are calculated as
a function of a static lattice distortion, in the spirit of the BO
principle, providing $\varphi({\bf r};[u])$. The frozen phonon
lattice potential is therefore used to solve the lattice quantum problem
and to get the ground state lattice wave function
$\chi({\bf u})$.
We shall now apply the same technique to MgB$_2$, in order to
show that, due to the small $E_{\rm F}^\sigma$
and to the strong deformation potential of the $E_{2g}$ phonon mode, the
BO approximation cannot be safely applied in this system.

As mentioned above,
small Fermi energy effects in MgB$_2$ involve
the subset of the almost two-dimensional
$\sigma$ bands with a corresponding Fermi energy
$E_{\rm F}^\sigma =$ 0.4-0.6 eV,\cite{an} defined as the distance in energy
between the chemical potential and the edge of the $\sigma$ band.
Furthermore, 
the electron-phonon interaction is essentially
dominated by the doubly-degenerate 
$E_{2g}$ phonon mode at ${\bf q}=0$,\cite{an,kong}
which gives
rise to a significant $\sigma$-$\sigma$ band electron-phonon
coupling, while the residual $\sigma$-$\pi$ and $\pi$-$\pi$
electron-phonon interaction is much weaker.\cite{liu,choi}
The effective Hamiltonian
of the $\sigma$ bands interacting with the $E_{2g}$ phonon mode
at ${\bf q}=0$ can be thus written as:\cite{bbcp}
\begin{equation}
H = \sum_{{\bf k}}
\epsilon_{{\bf k}}
c_{{\bf k}}^\dagger c_{{\bf k}}
\pm I_{E_{2g}} 
\sum_{{\bf k}}
c_{{\bf k}}^\dagger c_{{\bf k}} u
-\frac{\hbar^2 \nabla^2_u}{2M_{E_{2g}}}+V(u),
\label{ham}
\end{equation}
where $u$ is the lattice displacement of the $E_{2g}$ phonon mode
and the sign $\pm$ in the Jahn-Teller electron-phonon interaction
is different for the two $\sigma$ bands.
Assuming for the moment
a perfectly harmonic phonon mode with elastic constant 
$a_2$: $V(u) = a_2 u^2$, we can write Eq. (\ref{ham})
in second quantization [$u = (\hbar^2/4M_{E_{2g}}a_2)^{1/4}(a+a^\dagger)$]:
\begin{equation}
H = \sum_{{\bf k}}
\epsilon_{{\bf k}}
c_{{\bf k}}^\dagger c_{{\bf k}}
\pm g_{E_{2g}} 
\sum_{{\bf k}}
c_{{\bf k}}^\dagger c_{{\bf k}} (a+a^\dagger)
+\omega_{E_{2g}} a^\dagger a,
\label{ham2}
\end{equation}
where the electron-phonon matrix element ,
$g_{E_{2g}} = I_{E_{2g}}(\hbar^2/4Ma_2)^{1/4}$, determines the
change in the electron energy bands associated with
the ground state zero point motion.

Frozen phonon calculations have been previously employed
in Ref. \onlinecite{bbcp} to point out the connection between
small Fermi energy and
anharmonic effects of the $E_{2g}$ phonon mode in MgB$_2$.  The
appearance of anharmonic terms in the {\em static} phonon potential
has indeed been related to a strong lattice displacement regime,
namely: $D_{E_{2g}}^\sigma = I_{E_{2g}} u 
\bsim E_{\rm F}^\sigma$.
On physical grounds, we can expect that anharmonic terms
will be experimentally observable only if the lattice fluctuations
allowed in the system are sufficiently strong to sample regions of
phase space in which the effective Fermi energy is small 
($I_{E_{2g}} u \sim E_{\rm F}^\sigma$).  We are going to show that a
similar condition rules the breakdown of the adiabatic
BO principle.\cite{note}

\begin{figure}[t]
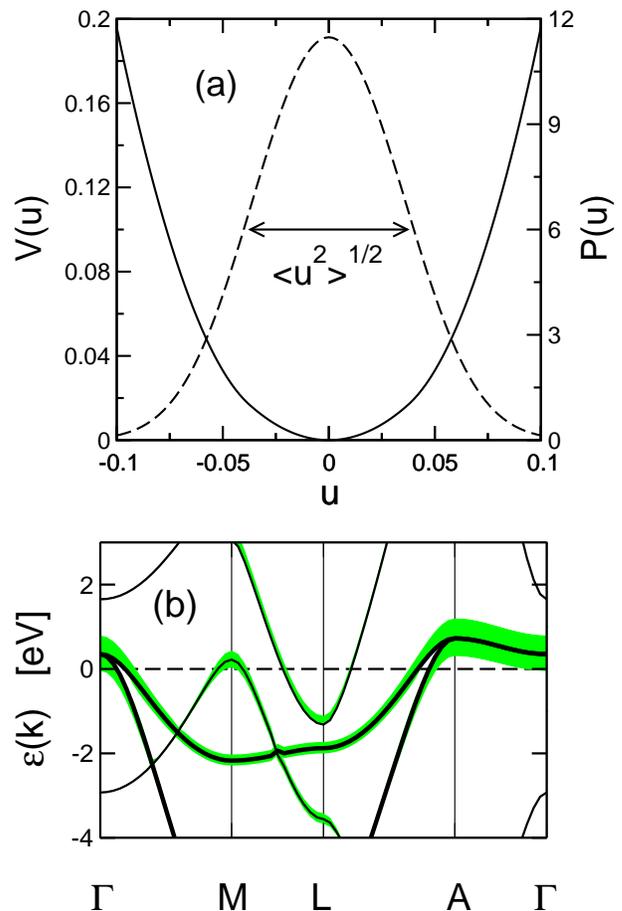

\centerline{\psfig{figure=well3.eps,width=8cm,clip=}}
\vspace{4mm}
\centerline{\psfig{figure=band-b.eps,width=8cm,clip=}}
\caption{(a) Frozen phonon potential $V(u)$ for the $E_{2g}$ mode
in MgB$_2$ (solid line, left side scale), and probability distribution
function $P(u)=|\psi_{E_{2g}}(u)|^2$ (dashed line, right side scale).
$u$ is in units of \AA, $V(u)$ in
units of eV and $P(u)$ in units of \AA$^{-1}$.
(b) Electronic structure of
undistorted MgB$_2$ (solid lines). The grey regions represent the
quantum fluctuations induced by zero point lattice motion ($|u|
\le \langle u^2 \rangle^{1/2}$). Bold lines represent the two
$\sigma$ bands, which are also the most affected by quantum
fluctuations.}
\label{lda-mgb2}
\end{figure}

In Fig. \ref{lda-mgb2}(a) we plot the $E_{2g}$ frozen phonon
potential $V(u)$ obtained by local density (LDA) calculations (see
Ref.  \onlinecite{bbcp} for details).  In the spirit of the BO principle we
can evaluate the ground state phonon wave function $\chi_{E_{2g}}(u)$
by the numerical solution of the Schr\"odinger equation
$[-(\hbar^2/2M_{E_{2g}})\nabla^2_u + V(u)] \chi_{E_{2g}}(u) = E_{E_{2g}}
\chi_{E_{2g}}$.  We can also define a lattice probability distribution
functions (PDF) $P(u)$ through the relations $P(u)=
|\chi_{E_{2g}}(u)|^2$, which is plotted in Fig. \ref{lda-mgb2}(a). 
In the classical (strictly adiabatic) $M_{E_{2g}} \rightarrow \infty$
limit, and $P(u)$ becomes a $\delta$-function centered at $u=0$.  The
electronic structure of the undistorted lattice, which corresponds
thus to the adiabatic limit, is shown in panel (b) as solid lines.

As mentioned above,
there are two different sources of quantum fluctuations of the lattice 
configuration with respect to the classical, static limit. 
The first one is given by the possibility
of quantum transitions between different electron
and lattice eigenfunctions.
This gives rise, when expanded around the lattice equilibrium
position $u=0$, to the conventional
electron-phonon scattering
which leads to the {\em  dynamical} electron-phonon renormalization
of the electronic and lattice properties, usually
taken into account by the corresponding self-energies.
A different effect of the quantum lattice fluctuations
is the fact that the ground state PDF
acquires a non-zero amplitude at $u \ne 0$, corresponding to the
zero-point motion. In this framework an appropriate quantity to
estimate the amplitude of the lattice quantum fluctuations 
is the root mean square (rms)
lattice displacement:   
\begin{equation}
\label{umedio}
\langle u^2 \rangle^{1/2} = \left[ \int du \,u^2 P(u)
\right]^{1/2}.
\end{equation}
In the following we focus on the interesting implications
of this second effect on the electronic structure.
We shall show that
the zero-point lattice fluctuations
are unavoidably
reflected, through the deformation potential, in an intrinsic
source of quantum fluctuations for the electronic structure and
for the Fermi energy.

A rough estimate of these effects is provided by their rms values. 
Applying the definition (\ref{umedio}), we obtain for the zero-point
motion of the $E_{2g}$ phonon mode a vibrational amplitude
$\langle u^2 \rangle^{1/2} \simeq 0.034$ \AA. 
The
changes in the electronic structure corresponding to this lattice
fluctuations are represented in Fig. \ref{lda-mgb2}(b) by the grey
regions. The main changes of the electronic structure close to the
$\Gamma$ point, $D_{E_{2g}}^\sigma = I_{E_{2g}} \langle u^2\rangle^{1/2}$  
can be obtained from the
deformation potential 
and roughly correspond to the thickness of the
grey region at the $\Gamma$ point. As shown in the figure, the
strong quantum {\em lattice} fluctuations are reflected in strong
{\em electronic} changes with a rms value of the $\sigma$ band
Fermi energy fluctuations
$\langle [ E_{\rm F}^\sigma (u) - E_{\rm F}^\sigma (0)]^2 \rangle^{1/2} 
= I_{E_{2g}} \langle u^2\rangle^{1/2}
\simeq 0.39$ eV, which is slightly
smaller than the electron-phonon matrix element $g_{E_{2g}}=0.45$ eV
because it includes anharmonic effects.
This value is 
comparable to the Fermi energy of the undistorted (adiabatic)
case $E_{\rm F}^\sigma = 0.45$ eV, resulting in 
a range of Fermi energy fluctuation 
$0.06 \mbox{ eV} < E_{\rm F}^\sigma < 0.84$ eV,
and casts doubts on the usual definition of the Fermi energy,
based on the BO approximation.

The relevance of this regime may be measured by the dimensionless parameter:
\begin{equation}
\label{kappa}
\kappa = \frac{\langle \left[ E_{\rm F}^\sigma (u) - E_{\rm
F}^\sigma (0) \right]^2 \rangle^{1/2}}{E_{\rm F}^\sigma} \simeq
\frac{I_{E_{2g}} \langle u^2 \rangle^{1/2}}{E_{\rm F}^\sigma}
\simeq 0.91.
\end{equation}
In common metals, although the numerator can be of the order of
a fraction of eV, the denominator is usually of $5-10$ eV, so
that $\kappa \ll 1$ and the role of the quantum lattice
fluctuations on the electronic structure is negligible. Things are
radically different for the $E_{2g}$ phonon of
MgB$_2$, where the large value of $\kappa$
($\simeq 1$) is driven by the extremely small Fermi energy of the
$\sigma$ bands.

The sizable magnitude of the parameter $\kappa$ calls for some
physical considerations.
Due to these strong quantum fluctuations, the
system indeed samples, with a sizable weight, electronic configurations with
zero or vanishing Fermi energy. The system has thus an intrinsic
nonadiabatic character \cite{ccgps} even if the phonon frequencies
$\omega_{\rm ph}$ are sensibly smaller than $E_{\rm F}(u=0)$. 
An alternative way to understand the breakdown of the BO principle is
to observe that for $u \simeq E_{\rm F}^\sigma / I_{E_{2g}}$, 
the electronic Fermi energy $E_{\rm F}^\sigma(u)$, 
which governs the 
dynamics of 
the $\sigma$ bands, is comparable with $\omega_{\rm ph}$,
which governs the 
lattice dynamics. The variable $u$ cannot thus be treated as
a classical degree of freedom in a parametric way and
$\Psi({\bf r};{\bf
u}) \neq \chi({\bf u}) \varphi({\bf r};[u])$.

We would like to stress once more that the origin of the nonadiabatic
breakdown of the BO principle
stems from the strong lattice quantum fluctuations
which make the electron-phonon matrix element,
the {\em second} term in Eqs. (\ref{ham})-(\ref{ham2}),
comparable with the Fermi energy.
This is thus qualitatively {\em different}
from what discussed in Refs. \onlinecite{ccgps,ccgp},
where the nonadiabatic effects are ruled by the ratio between
the energy scale of the {\em third} term
in Eq. (\ref{ham2}) and the Fermi energy,
namely $\omega_{E_{2g}}/E_{\rm F}^\sigma$.
Note also that
the phonon frequency $\omega_{E_{2g}}$
does {\em not} depend on the amplitude
of the lattice fluctuations, on the contrary of the
parameter $I_{E_{2g}} \langle u^2 \rangle^{1/2} /E_{\rm F}^\sigma$.
This means that the onset of one kind of nonadiabatic effects
does not necessarily implies the onset of the
other one, and {\em vice versa}.

In order to further clarify this statement let us assume for the moment
a perfectly harmonic phonon mode with elastic constant 
$a_2$: $V(u) = a_2 u^2$.
Anharmonic contributions do not play any important role 
in the following discussion, so we will  take them into account through
an ``effective'' elastic constant as in Ref. \onlinecite{yildirim}
to give the anharmonic hardening of the phonon.

\begin{figure}[t]
\centerline{\psfig{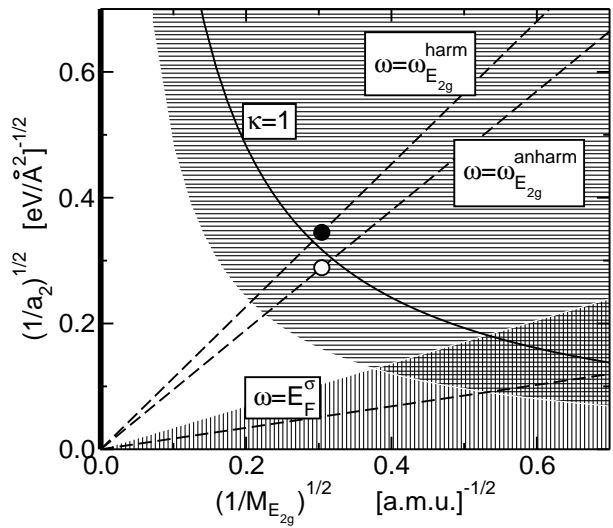}}
\caption{Schematic phase diagram for the different kinds of
nonadiabatic effects. See text for details.}
\label{fig-ph}
\end{figure}

The two different sources of breakdown of the BO principle 
are illustrated in Fig. \ref{fig-ph} where we plot
on the $x$ axis the
inverse of the square root of the phonon mass $1/\sqrt{M_{E_{2g}}}$ and
on the $y$ axis
the inverse of the square root of
the elastic constant $1/\sqrt{a_2}$. 
The limit $M_{E_{2g}} \rightarrow \infty$, denoted by the thick line,
is the strictly classical (adiabatic) case where both kinds of
nonadiabatic effects are negligible.
In the so-defined phase space a fixed phonon frequency is given
by a straight line cutting through the origin.
In Fig. \ref{fig-ph} we show as dashed lines the three frequencies 
$\omega_{E_{2g}}=68$ meV, $81$ meV, and $0.45$ eV,
corresponding
to the harmonic and anharmonic frequencies of the $E_{2g}$ phonon, 
and to a hypothetical phonon in the fully nonadiabatic regime
($\omega_{\rm ph}=E_{\rm F}^\sigma$) respectively.
The empty (filled) circles represent respectively
the actual position of the $E_{2g}$ phonon of MgB$_2$
($M=10.81$ a.m.u.)
with (without) taking into account
the anharmonic hardening.
The nonadiabatic breakdown of the BO principle underlined in
Ref. \onlinecite{ccgps} arises in the region
$\omega_{\rm ph} \simeq E_{\rm F}^\sigma$,
and is represented by the vertically-dashed area,
corresponding to 
$\omega_{\rm ph}/ E_{\rm F}^\sigma \ge 0.5$.

The breakdown of the BO principle due to the zero point motion, determined
by the parameter $\kappa$,
of the lattice fluctuation is also shown in Fig. \ref{fig-ph}.
$\kappa$-isolines are easily defined by the relations
$a_2 \langle u^2 \rangle^{1/2} = \hbar \omega_{\rm ph} /2$ and
$\kappa = I_{E_{2g}} \langle u^2 \rangle^{1/2} /  E_{\rm F}$ using
LDA values $I_{E_{2g}} = 12$ eV / \AA\, and 
$E_{\rm F}^\sigma = 0.45$ eV. The
horizontally-dashed region ($\kappa \ge 0.5$) represents the
range of parameters where nonadiabatic effects could be triggered by the
zero-point fluctuations. It is interesting to note that
in this new regime nonadiabatic effects induced by 
lattice fluctuations
can be operative even quite far from the usual 
``nonadiabatic'' regime 
$\omega_{\rm ph} \sim E_{\rm F}$. Of course, the two sources of
nonadiabaticity merge together in the highly nonadiabatic case
$M \rightarrow 0$, where $E_{\rm F}$ is
smaller than $\omega_{\rm ph}$ already in the absence of fluctuations.
This could be the case for the cuprates and fullerene compounds
since in their case $E_{\rm F}$
is smaller than in MgB$_2$, while the relevant phonon frequencies are 
comparable.

In conclusion, in this work we have pointed out
the inadequacy in MgB$_2$ of the BO approximation
which is at the basis of many {\em ab-initio} techniques,
when applied to coupling 
between the $E_{2g}$ phonon and the $\sigma$ holes, which are
mainly responsible for superconductivity. We have related
such inadequacy to the strong
lattice fluctuations of zero-point motion which,
due to the large deformation potential associated with the $E_{2g}$
phonon, induce electronic changes of the $\sigma$ bands
of the same order of magnitude
of their Fermi energy ($E^{\sigma}_{\rm F} \simeq 0.45$ eV).
The breakdown of the BO approximation has important consequences
since it implies the failure of 
many {\em ab-initio} techniques.
An appropriate inclusion of these effects is a
formidable task which
goes beyond the aim of this paper.
We note however
that the structural and cohesive properties
of MgB$_2$ are expected to be weakly affected by the zero-point
fluctuations, both because they depend on the whole set of occupied
bands (whose width is 10-20 eV), and because they involve all
phonons. Its transport and
superconducting properties, instead, are mainly determined from the
small fraction of carriers within an energy
range $\sim \omega_{\rm ph}$ close to the
Fermi level, and which are strongly coupled to the relevant $E_{2g}$ phonon.

As a last remark, it is interesting to remind that
the nonadiabatic effects induced by
lattice fluctuations are increased as the force constant decreases.
Electron-phonon systems with incipient lattice instabilities
(corresponding to flattening of the lattice potential)
could be thus good candidates for the observation of similar nonadiabatic
effects in other materials than MgB$_2$.
Good candidates in this direction are bismuthates, A15 compounds
and possibly monoatomic metals as Nb or V.
In particular, the important role of the lattice fluctuations in A15
compounds, which was already pointed out by P.W. Anderson
in Ref. \onlinecite{anderson}, could provide the basis to investigate some
interesting similarities between A15 and diborides systems.

We thank G.B. Bachelet, C. Grimaldi, S. Ciuchi and M. Capone
for many fruitful discussions.
We acknowledge also the project PRA-UMBRA of the INFM
for financial support.


\begin{thebibliography}{99}
\bibitem{note}
In the following thermal fluctuations will be safely disregarded
since they are not expected to play any role as long as $T \ll
E_{\rm F}^\sigma \simeq 5000$ K.
\end{thebibliography}
\end{document}